\documentclass[aps,prd,twocolumn,superscriptaddress,nofootinbib,preprintnumbers,longbibliography]{revtex4-2}

\usepackage[utf8]{inputenc}
\usepackage{amsmath,amssymb}
\usepackage{hyperref}
\usepackage{mathrsfs}
\usepackage{bm}
\usepackage{graphicx}
\usepackage{epsfig}

\usepackage[usenames,dvipsnames]{xcolor}
\definecolor{orange}{cmyk}{0,0.5,1,0}
\definecolor{rossoCP3}{cmyk}{0,.88,.77,.40}
\definecolor{graa}{rgb}{0.8,0.8,0.8}
\definecolor{blaa}{rgb}{0.2,0.2,0.6}

\begin{document}

\title{Cosmology \& Astrophysics of the Dark Dimension with Bulk Majorana Neutrinos}

\author{\bf Luis A. Anchordoqui}

\affiliation{Department of Physics and Astronomy,  Lehman College, City University of
  New York, NY 10468, USA
}

\affiliation{Department of Physics, 
 Graduate Center,  City University of
  New York,  NY 10016, USA
}

\affiliation{Department of Astrophysics,
 American Museum of Natural History, NY
 10024, USA
}

\author{\bf Cumrun Vafa}

\affiliation{Jefferson Physical Laboratory, Harvard University, Cambridge, MA 02138, USA}

\begin{abstract}
\noindent Recently, the possibility of having sterile Majorana neutrinos charged under $B-L$ gauge symmetry in the bulk of the dark dimension has been explored.
We point out a cosmological restriction on the Majorana mass of bulk neutrinos  leading to mass $\gtrsim 100~{\rm GeV}$.  This avoids the over-production of Kaluza-Klein neutrino towers in the early universe thus resolving the cosmological neutrino catastrophe of extra dimension scenarios.   If the mass of the bulk neutrino is at the lower end of this bound, it can potentially also explain the recently observed Fermi halo excess on gamma rays through the evaporation of long-lived 5d (i.e. dark) black holes through Hawking radiation to bulk neutrinos, which promptly decay into standard model particles. 
  \vskip 2mm
  % \noindent
\end{abstract}

\maketitle

\newpage

\section{Introduction}

The dark dimension introduces a compelling framework
where our universe contains a single mesoscopic extra
dimension of roughly micron size,
offering a unified description of
 dark energy and dark
 matter~\cite{Montero:2022prj,Gonzalo:2022jac}. 
 The universal coupling
 of standard model (SM) fields localized on a brane to bulk
 gravitons inevitably creates a tower of Kaluza-Klein (KK)
 modes in the early universe. These massive spin-2 dark gravitons possess a
uniform mass splitting ($m_{\rm KK} \sim 0.1~{\rm
eV}$) dictated by the micron scale $R$ of the dark dimension. Because
the production of a specific mode $m_n \sim n m_{\rm KK}$ is highly
dominant at a temperature matching its own mass ($T_n \sim m_n$)
, the
early universe quickly becomes populated with a massive, 
ensemble of heavily excited modes.
In this model, the initial
temperature $T_*$ is when the dark graviton produced would constitute the observed dark matter abundance.  We also assume that at this temperature the early universe has already settled from
a higher-dimensional system into our
familiar, effectively
4d spacetime with stabilized geometric
moduli. To remain consistent with
observational constraint on the dark matter abundance
 the initial
temperature is approximately
$T_* \sim 1~{\rm GeV}$~\cite{Gonzalo:2022jac}, which also preserves the successes of Big Bang
nucleosynthesis since it is before it ($T_*>T_{\rm BBN}$).

The cosmological evolution of these dark gravitons can proceed either by
KK graviton intra-tower decays~\cite{Gonzalo:2022jac} or by conversion to KK tower of quintessence field~\cite{Vafa:2026ccm} which would be more natural for connection with DESI data~\cite{Bedroya:2025fwh}. The KK momentum of dark dimension is not conserved.
This allows  a heavily excited, KK tower
fragment by decaying into two lighter KK states  within the
tower. This intra-tower cascade provides a specific realization of the dynamical dark
matter framework, continuosly shuffling the energy density down the
tower toward lighter, increasingly stable modes~\cite{Dienes:2011ja}. To avoid radiative decay signatures~\cite{Law-Smith:2023czn} and the
premature disruption of bound structures~\cite{Obied:2023clp}, a
dynamical dark sector governed by KK tower cascades requires a
fast enough evolving tower with a time-dependent mass scale of $m_n(t) \propto t^{-2/7}$.\footnote{Two derivative restriction of the 5d effective theory can lead to suppression of the KK
  graviton cascade rate, slowing down the mass evolution to $m_n(t)
  \propto t^{-2/3}$~\cite{Lee:2026myr,Langhoff:2026wiz} which can be avoided either by allowing higher derivative terms or more naturally by allowing gravitons to promptly decay into the quintessence field  (and the subsequent scalar tower cascade preserves the
 $t^{-2/7}$ scaling) leaving the cosmological evolution of \cite{Gonzalo:2022jac}
 intact~\cite{Vafa:2026ccm}.} Of particular interest to our
discussion is the survival of dwarf satelite galaxies and low-mass
subhalos. Because dark-to-dark decay happens continuously, the momentum kicks
inject kinetic energy into existing structures. If the average kick velocities exceed $v_{\rm  kick} \sim 10^{-4} c$, the injected kinetic energy reduces the central density of cuspy
dark matter profiles, evaporating the outer envelopes of galactic
structures and depleting low-mass subhalos via enhanced tidal
stripping. Therefore, to satisfy structural constraints, the decay of
the $n$-th KK mode must occur across highly compressed mass gaps where the
mass splitting $\Delta m \ll m_n$. This heavily restricts the phase
space, ensuring low-momentum violation. The kinematics of the recoil
process dictate that the daughter state receives a kick velocity given
in the non-relativistic limit by $
v_{\rm kick} \simeq c \, \Delta m/m_n$.

Beyond serving as a framework for connecting dark energy and dark matter, this
geometric structure provides a natural
gateway for understanding neutrino
masses and mixing through
bulk physics especially assuming a $B-L$ bulk gauge symmetry \cite{Montero:2025hye} (sterile neutrinos in the bulk were also considered in the context of large extra dimensions~\cite{Dienes:1998sb,Arkani-Hamed:1998wuz, Dvali:1999cn}). Because right-handed neutrinos carry no gauge charges under the SM, they are free to propagate throughout the 5d bulk. The coupling between these bulk states and the brane-localized, left-handed SM neutrinos scales inversely with the square root of the length of dark dimension.

The oscillation of active SM neutrinos into sterile KK modes offers a
unique experimental signature for probing the 
brane-bulk neutrino mixing~\cite{Davoudiasl:2002fq}. More concretely, this signature manifests as a distinct
disappearance pattern when active neutrinos oscillate out of our 4d spacetime into the bulk. Because this leakage into the dark dimension
alters standard neutrino survival probabilities in a predictable way, its presence can be rigorously
tested. The
  geometrically suppressed brane-bulk neutrino mixing framework can
  also be probed by scanning the high-precision electron energy spectrum
  of tritium beta decay for its unique neutrino signal, looking
  specifically for distinct KK mass kinks near the endpoint energy~\cite{Antoniadis:2025rck}. To date, short- and long-baseline neutrino oscillation experiments have failed to detect this characteristic disappearance.
Consequently, these null results have been used to set bounds on the size of the extra dimension~\cite{Machado:2011jt,Forero:2022skg}. Notably, these bounds could be significantly relaxed in the
presence of bulk neutrino masses~\cite{Carena:2017qhd,Anchordoqui:2023wkm,Bai:2026kdq}.

Despite these viable detection pathways, the fundamental bottleneck
facing the
brane-bulk neutrino mixing framework is the cosmological neutrino catastrophe, a severe
overproduction of KK modes in the early
universe in the absence of masses for the bulk sterile neutrinos~\cite{Barbieri:2000mg,Abazajian:2000hw}. This runaway production introduces a multi-pronged cosmological threat: the production 
of exceptionally dense light KK modes drastically injects extra relativistic degrees of freedom 
that violate strict observational limits on the effective number of
neutrino species ($N_{\text{eff}}$), whereas the population of the distribution's tail with unsuppressed heavy KK modes carries 
enough relic energy density to completely overclose the universe. In this note, we examine 
whether introducing explicit bulk neutrino masses can alter the active-sterile mixing profile, 
providing a unified active-sterile suppression mechanism capable of simultaneously quenching 
both the light relativistic excitations and the heavy overclosing states to reconcile the 
framework with global cosmological bounds.\footnote{A point worth
  noting at this juncture is that secret
  neutrino interactions via light mediators can suppress
  active-sterile mixing at high temperatures, effectively blocking
  early-universe sterile neutrino production while preserving vacuum
  mixing today~\cite{Hannestad:2013ana,Dasgupta:2013zpn}. However, the
  allowed parameter space for these secret interactions is severely
  constrained by cosmological bounds on active neutrino free-streaming
  and the sum of neutrino masses~\cite{Chu:2015ipa,Chu:2018gxk}.}

We work within the construct of~\cite{Montero:2025hye}
 featuring a $B-L$ gauge symmetry that naturally propagates into the
 bulk, inducing an exact global symmetry on the brane while charged
 bulk modes self-consistently cancel gauge anomalies.  In the dark dimension scenario, $B-L$ can also naturally be a gauge symmetry on the SM brane broken by GUT physics~\cite{Heckman:2024trz}, but here we consider the other possibility of a dark $B-L$ gauge symmetry in the bulk. This geometric
 setup naturally introduces three right-handed neutrinos propagating
 through the dark dimension. When the $B-L$ charged bulk scalar develops
 a vacuum expectation value to trigger symmetry breaking it yields a massive gauge boson
 with an exceptionally feeble coupling ($g_{B-L} \sim 10^{-10}$),
 remaining fully consistent with existing observational
 bounds. Notably, the right-handed neutrinos couple to the bulk scalar, which  generates a Majorana mass upon Higgsing. By construction, the
compactification of this dark dimension dictates the emergence of a KK
tower of sterile right-handed neutrinos. We investigate the cosmological consequences of exciting these three
distinct neutrino towers. We
demonstrate that a Majorana mass larger than 100 GeV scale suppresses the
excitation of neutrino ladders in the early universe, leaving the late
time cosmology to be that of the model proposed in~\cite{Gonzalo:2022jac}. Under this baseline scenario, dark matter is comprised entirely of the KK tower(s) as in \cite{Gonzalo:2022jac}.

The remainder of this work is structured as follows. We begin in
Sec.~\ref{sec:2} by reviewing the main characteristics of the
neutrino sector introduced in~\cite{Montero:2025hye}, detailing the
underlying setup. In Sec.~\ref{sec:3}, we analyze the cosmological
implications of this set up and show how to prevent the neutrino catastrophe. Next, in Sec.~\ref{sec:4} we examine potential experimental tests of the model in the present-day universe. The paper wraps up in Sec.~\ref{sec:5} with some conclusions.

\section{The Dark Dimension Equipped with Majorana Neutrinos}
\label{sec:2}

To lay the groundwork for our subsequent analysis, in this section we
provide a concise overview of the framework introduced
in~\cite{Montero:2025hye}. Even though it is not necessary, for concreteness let us consider a 5d flat spacetime where the extra spatial dimension $y$ is
compactified on an $S^1/\mathbb{Z}_2$ orbifold of length $R\sim
1~\mu{\rm m}$~\cite{Montero:2022prj}. This configuration forces the bulk
fields to transform under the parity reflection $y \to -y$,
circumventing the absence of chirality inherent to standard circle
compactifications. The SM fields,
including the left-handed lepton doublet $L$ and the Higgs doublet
$H$, are confined to a brane located at $y = 0$, while gravity spills into the 5d bulk. Such a construction leads to
the well-known relation $M_p^2 \sim R M_5^3$, between the  4d ($M_p$) and 5d $(M_5)$ reduced Planck scales. We take $M_5\sim 10^9~{\rm GeV}$.

Next, we introduce the neutrino spectrum following the approach
outlined in~\cite{Montero:2025hye}. Although the SM comprises three
neutrino generations, to simplify the notation we restrict our
attention to a single flavor and omit the flavor index 
unless otherwise specified. Consider one active Weyl neutrino $\nu_L$ on the brane and a tower of sterile Majorana fermions $\psi_n$  with masses
\begin{equation}
  m_n \sim \sqrt{m_\psi^2 + n^2/R^2}  \, ,
\end{equation}
and $n \in \mathbb Z$. More concretely, the $n$-th KK level of
$\psi$ is structured as a pair of Weyl fermions characterized by a Dirac mass $n
m_{\rm KK}$ alongside a Majorana mass $m_\psi$. A brane-localized Yukawa couples $\nu_L$
to every parity-even Weyl fermion with the same strength
\begin{equation}
m_D \sim  \frac{\lambda_Y \langle H \rangle}{\sqrt{M_5 R}} \,,
\end{equation}
(up to $\sqrt{2}$'s of the orbifold) where $\langle H \rangle \approx
175~{\rm GeV}$ is the Higgs vacuum expectation value and $\lambda_Y$ the
Yukawa coupling. Specifically, in the basis $\{\nu_L, \psi_0^c, \psi_1^c, \psi_{-1}^c, \cdots\}$, the
mass matrix is given by
\begin{equation} \mathbb{M} = \left(\begin{array}{ccccccc} 0 & m_D
      &\widetilde m_D & \widetilde m_D & \dots & \widetilde m_D &
      \widetilde m_D \\ m_D & m_\psi
                                      & 0 & 0 & \dots & 0 & 0 \\
                                      \widetilde m_D & 0 & m_1 & 0 &
                                                                     \dots
                                               & 0 & 0 \\ \widetilde
                                      m_D & 0 & 0 & m_{-1} & \dots & 0
                                                                & 0 \\
                                      \vdots & \vdots & \vdots &
                                                                 \vdots
                                       & \ddots & \vdots & \vdots \\
                                      \widetilde m_D & 0 & 0 & 0 &
                                                                   \dots
                                               & m_N & 0 \\ \widetilde
                                      m_D & 0 & 0 & 0 & \dots & 0 &
                                                                    m_{-N} \end{array}\right)\,, \label{mass_matrix} \end{equation}
                                                                    with
                                                                    $\widetilde
                                                                    m_D
                                                                    =
                                                                    \sqrt{2}
                                                                    m_D$~\cite{deGiorgi:2026isr}.

                                                                    For
                                                                    $m_\psi
                                                                    \gg
                                                                    m_D$,
                                                                    integrating
                                                                    out
                                                                    the
                                                                    $n$-th
                                                                    level
                                                                    adds
                                                                    a
                                                                    contribution
                                                                    of
                                                                    $m_D^2
                                                                    m_\psi/(m_\psi^2
                                                                    +
                                                                    n^2
                                                                    m_{\rm
                                                                      KK}^2)$
                                                                    to
                                                                    the
                                                                    active
                                                                    neutrino's
                                                                    Majorana
                                                                    mass. Because
                                                                    every
                                                                    level
                                                                    in
                                                                    the
                                                                    tower
                                                                    share 
                                                                    the
                                                                    same
                                                                    sign,
                                                                    summing
                                                                    over
                                                                    all
                                                                    of
                                                                    them
                                                                    yields
                                                                    a
                                                                    compact
                                                                    analytic
                                                                    expression
                                                                    for
                                                                    the
                                                                    overall
                                                                    neutrino
                                                                    mass~\cite{deGiorgi:2026isr,Concha:2026}
\begin{eqnarray}
m_\nu & \simeq &  m_D^2   \sum_n \frac{m_\psi}{m_\psi^2 + n^2
                                                                 m_{\rm
                                                                   KK}^2}
                                                                        \nonumber \\
                                                              &
                                                                \approx &
                                                               \frac{\lambda_Y^2
                                                                 \langle
                                                                 H\rangle^2
                                                                          \pi}{M_5}
                                                               \coth
                                                               (\pi
                                                               m_\psi
                                                               R)
                                                                          \approx  \frac{\lambda_Y^2
                                                                 \langle
                                                                 H\rangle^2
                                                                          \pi}{M_5}
                                                                          \,  .
                                                             \label{mnuformula}             
\end{eqnarray}
The observed masses therefore fix  $\lambda_Y \sim 10^{-3}$, comparable to the
charm or $\tau$ Yukawa coupling, yielding $m_D = (m_\nu m_{\rm KK}/\pi)^{1/2}
\simeq 0.06~{\rm eV}$. The admixture of mode $n$ in the light state is found to be $\theta_n
= m_D/m_n$, and the total is UV-finite:
\begin{equation}
  \sum_n \theta_n^2 =  m_D^2 \sum_{n\in \mathbb Z} \frac{1}{m_\psi^2 +
  n^2 m_{\rm KK}^2} \simeq \frac{m_\nu}{m_\psi} \,.
\end{equation}
Although a large seesaw mass $m_\psi$ renders the active-sterile
mixing angle to a negligible level, its implications remain highly
relevant for both early-universe cosmology and local astrophsyics. It
is this that we now turn to study.

\section{Cosmological Implications of Neutrinos in the Dark Dimension}
\label{sec:3}

In this section, we establish a lower bound on the mass $m_\psi$ by evaluating
the constraints imposed by the initial temperature. We
consider a scenario where KK gravitons saturate the observed relic
density, with $T_* \sim 1~{\rm GeV}$~\cite{Gonzalo:2022jac}.  Under these conditions, it is crucial to prevent any additional,
subdominant contributions from KK neutrino ladders, which would otherwise overclose
the universe or conflict with standard cosmological evolution. By synthesizing these
boundary conditions and applying the relic density limits,
we map out the allowed parameter space for the sterile neutrino sector, thereby
placing a definitive lower limit on the mass gap $m_\psi$.

The effective conversion rate $\Gamma(\nu_\alpha \to n)$
of an active neutrino flavor $\nu _{\alpha }$ into the $n$-th KK mode via oscillation and subsequent active-neutrino scattering in the plasma is given by
\begin{equation}
\Gamma (\nu _{\alpha }\rightarrow n)=\frac{\Gamma_{\rm act}}{2} \ \langle
P(\nu _{\alpha }\rightarrow n)\rangle \, 
\end{equation}
where $\Gamma_{\rm act}$ is the active neutrino's total
SM plasma interaction rate and $\langle
P(\nu _{\alpha }\rightarrow n)\rangle$ 
the time-averaged
transition probability~\cite{Abazajian:2000hw}.

To evaluate the cosmological abundance of KK towers, it is 
necessary to account for how thermal plasma effects modify sterile neutrino 
production. In the early universe, background forward scattering dresses 
active neutrinos with a finite-temperature effective self-energy. This 
self-energy acts essentially as a refractive index for the active neutrino 
propagation, inducing a thermal potential $V_T$ that modifies its effective 
dispersion relation~\cite{Notzold:1987ik}. This medium-induced potential 
alters the forward-scattering amplitude, shifting the effective in-medium 
mixing angle away from its vacuum value and suppressing active-to-sterile 
neutrino oscillations at high temperatures~\cite{Dodelson:1993je}.

Now, for a relativistic neutrino in thermal equilibrium at temperature
$T \ll m_W$, the typical average momentum is $p \sim 3T$ and the
thermal potential scales as
\begin{equation}
|V_T| \sim G_{\rm eff} \, p \, T^4 \sim 10~{\rm eV} \left(\frac{p}{{\rm
      GeV}}\right) \, \left(\frac{T}{{\rm GeV}}\right)^4 \,,  
\end{equation}
where $m_W$ is the $W$-boson mass and $G_{\rm eff}$ can be taken to
represent some overall neutrino interaction strength, summed over all
the particle species in the background
plasma~\cite{Patwardhan:2015kga}. This in turn implies that for modes with $m_n \gtrsim 1~{\rm MeV}$ at a cosmic temperature of 
$T \sim 1~{\rm GeV}$, the effect of this thermal potential becomes entirely negligible 
because $|V_T|/\Delta \sim 0.2 ({\rm MeV}/m_n)^2$, where 
$\Delta \simeq m_n^2/(2p)$ is the characteristic vacuum oscillation 
frequency within the thermal bath. Consequently, the refractive effects of 
the plasma fade for these heavy states, allowing every KK mode lighter than 
$T_*$ to escape plasma suppression and populate the bulk via its standard 
vacuum mixing angle. Assuming then a pure vacuum unaffected by plasma, the time-averaged transition probability to a single mode $n$ reduces to
\begin{equation}
\langle P(\nu _{\alpha }\rightarrow n)\rangle \simeq
\frac{1}{2} \sin^2 (2\theta_{n}) \, .
\end{equation}
Given that the standard active neutrino interaction rate scales as $\Gamma_a \simeq 2.9 G_F^2 T^5$~\cite{Enqvist:1991qj}, the mode-specific production rate becomes:
\begin{equation}
  \Gamma \simeq 2.9 \theta_n^2 \, G_F^2 \, T^5 \, ,
  \label{G7}
\end{equation}
where $G_F$ is the Fermi coupling constant.

The present-day comoving yield of the $n$-th KK mode normalized to the
total entropy density $s_{\rm today}$
is given by
\begin{equation}
Y_{n,{\rm today}} \equiv \frac{n_{n,{\rm today}}}{s_{\rm today}} \,,
\end{equation}
where $n_{n,{\rm today}}$ is its number density.  Because the total entropy in a comoving volume ($S = s \cdot a^3$) is
conserved, this yield ratio remains invariant once the production of
the KK tower freezes out at the temperature $T_* \sim 1~{\rm
  GeV}$. Thus, we can scale out the active neutrino background density
at freeze-out ($n_{\nu *}$) to evaluate $Y_{n,{\rm today}}$,
\begin{equation}
Y_{n,{\rm today}} = Y_n(T_*) = \frac{n_n(T_*)}{s(T_*)} = 
  \frac{n_n(T_*)}{n_{\nu *}}\, \frac{n_{\nu
      *}}{s(T_*)} .
\label{Yntoday}
\end{equation}
For an ultra-relativistic active neutrino species (accounting for both particles and antiparticles per flavor), the number density is
\begin{equation}
n_{\nu *} = \frac{3 \zeta(3)}{4\pi^2} g_\nu T_*^3
\end{equation}
where $g_\nu = 2$. The total ambient entropy density is defined via the effective relativistic degrees of freedom $g_{{\rm eff}}$
\begin{equation}
s(T_*) = \frac{2\pi^2}{45} g_{{\rm eff}}(T_*) T_*^3
\end{equation}
Setting $g_{{\rm eff}}(T_*) \approx 60$ at the GeV scale, the background
scaling factor of (\ref{Yntoday}) evaluates numerically to
\begin{equation}
\frac{n_{\nu *}}{s(T_*)} \simeq 0.02 \, ,
\end{equation}
where we have considered 3 neutrino flavors.

In a spatially flat, expanding Friedmann--Lema\^{i}tre--Robertson--Walker universe, the un-backreacted Boltzmann equation tracking the number density $n_n$ of the sterile mode is
\begin{equation}
\label{boltzmann_nl}
\frac{dn_n}{dt} + 3Hn_n \simeq \Gamma n_{\nu}
\end{equation}
where $H$ is the Hubble expansion parameter and $n_{\nu}$ is the background active neutrino number density~\cite{Dodelson:1993je}. Crucially, the active background neutrino density $n_\nu$ also dilutes due to the expansion,
\begin{equation}
\label{boltzmann_nnu}
\frac{dn_\nu}{dt} + 3Hn_\nu = 0 \implies \frac{dn_\nu}{dt} = -3Hn_\nu
\, .
\end{equation}
Now, to eliminate the expansion dilution term, we transform variables to track the abundance ratio $n_n / n_{\nu}$
\begin{equation}
\frac{d}{dt} \left( \frac{n_n}{n_\nu} \right) = \frac{1}{n_\nu}
\frac{dn_n}{dt} - \frac{n_n}{n_\nu^2} \frac{dn_\nu}{dt} \, .
\label{tran}
\end{equation}
Substituting (\ref{boltzmann_nl}) and (\ref{boltzmann_nnu}) into (\ref{tran})
leads to
\begin{equation}
\frac{d}{dt} \left( \frac{n_n}{n_\nu} \right) = \Gamma \, .
\end{equation}
During the radiation-dominated era, $d/dt = - H T d/dT$, yielding
\begin{equation}
\frac{d}{dT} \left( \frac{n_l}{n_{\nu}} \right) = - \frac{\Gamma}{H T}
\, ,
\end{equation}
where the Hubble parameter scales as
\begin{equation}
  H(T) \simeq 0.33 \sqrt{g_{\rm eff}} \frac{T^2}{M_p} \, .
\label{HT}
\end{equation}
Combining (\ref{G7}) and (\ref{HT}) results in
\begin{equation}
\frac{d}{dT} \left( \frac{n_l}{n_{\nu}} \right) =  - 9 \frac{
  \theta_l^2 \, G_F^2 \, M_p}{\sqrt{g_{\rm eff}}} \,  T^2 \, .
\end{equation}
Integrating we arrive at the cumulative ratio
\begin{equation}
\frac{n_n}{n_{\nu*}} \simeq 9 \int_0^{T_*} \frac{\theta_n^2 G_F^2
  M_p}{\sqrt{g_{\rm eff}}} T^2 \, dT = 3 \frac{\theta_n^2  G_F^2 M_p
}{\sqrt{g_{\rm eff}}} T_*^3 \, 
\end{equation}
and so
\begin{equation}
  Y_{n,{\rm today}} \sim 2.5 \times 10^6 \ \theta_n^2 \ \left(\frac{T_*^3}{{\rm
        GeV}}\right)  \ \bigg(\frac{g_{\rm eff}}{60} \bigg)^{-1/2} \, .
\end{equation}  

All in all, the total energy density of the neutrino tower today is estimated
to be
\begin{equation}
\rho^{{\rm tot}}_{{\rm today}} = \sum_{n} m_n n_{n, {\rm today}} =
s_{\rm today} \sum_{n} m_n Y_{n, {\rm today}} \,,
\end{equation}
and so the cosmological parameter of the tower evaluates to
\begin{eqnarray}
\Omega_{{\rm tower}} h^2 & = & \frac{s_{\rm today}}{\rho_{c,
                               {\rm today}}/h^2}
                               \ \sum_{n} m_n Y_{n, {\rm today}} \nonumber
  \\
  & \simeq & 2.9 \times 10^8~{\rm GeV}^{-1} \  \sum_{n}
             m_n \ Y_{n, {\rm today}} \, ,
\end{eqnarray}
where  $\rho_{c, {\rm today}} \approx 1.05 \times 10^{-5} h^2 \text{ GeV cm}^{-3}$
is the critical density, with $s_{\rm today} \simeq 3000~{\rm
  cm}^{-3}$ and $h$ the dimensionless Hubble constant.

Next, we calculate
\begin{equation}
\sum_n \theta_n^2 m_n = \sum_n \frac{m_D^2}{m_n}
\end{equation}
for $m_n < T_*$. When the modes are densely spaced relative to the high-temperature cutoff scale ($T_* \gg m_{\rm KK}$), the discrete sum over the index $n \ge 1$ is approximated as a continuous integral,
\begin{equation}
\sum_n \frac{1}{m_n} \;\longrightarrow\;2 \int_{0}^{n_{\rm max}} \frac{dn}{\sqrt{m_\psi^2 + n^2 m_{\rm KK}^2}}
\end{equation}
where $n_{\rm max} \simeq T_*/m_{\rm KK}$. Evaluating the integral we obtain
\begin{equation}
 \int_{0}^{n_{\rm max}} \frac{dn}{\sqrt{m_\psi^2 + n^2 m_{\rm
      KK}^2}} = \frac{1}{m_{\rm KK}}
\ {\rm arcsinh}\left(\frac{T_*}{m_\psi}\right) \, ,
\end{equation}
and so
\begin{equation}
\sum_n  \theta_n^2 m_n \simeq \frac{2}{\pi} m_\nu \ {\rm arcsinh}
    \left(\frac{T_*}{m_\psi} \right)\, .
\end{equation}    

For heavy modes ($m_n > T_*$), the production rate experiences a
kinematic Boltzmann-like suppression of the form $(m_n/T_*)^2 e^{1 - m_n/T_*}$. To determine the cumulative weight of the $1/m_n$ factor across the
kinematically suppressed tower, setting $dn = dm_n/m_{\rm KK}$ we evaluate the integral 
\begin{equation}
\int_{m_\psi}^{\infty} \frac{m_n}{T_*^2} \,
e^{1 - \frac{m_n}{T_*}} \, dm_n = \left(1 + \frac{m_\psi}{T_*}\right)
e^{1 - \frac{m_\psi}{T_*}} \, .
\end{equation}

Finally, putting all this together we arrive at
\begin{equation}
\Omega_{\rm tower} h^2 \sim {\cal A} 
\times \left\{\begin{array}{ll} {\rm arcsinh}(T_*/m_\psi), &
                                                             m_\psi<T_*
                \\
                (1+ m_\psi/T_*) e^{1-m_\psi/T_*}, & m_\psi>
                                                   T_* \end{array}
                                               \right ., \nonumber
                                               \end{equation}
                                               \begin{equation}
                                                 {\rm with} \quad \quad {\cal A}
                                                 \sim 5 \times 10^4
                                                 \left(\frac{T_*}{{\rm
                                                       GeV}}\right)^3
                                                 \bigg(\frac{g_{\rm eff}}{60}\bigg)^{-1/2}
                                                 \, .
\label{calA}
\end{equation}                                                 
Now, a simple calculation using (\ref{calA}) demonstrates that for $m_\psi \lesssim T_* \sim 1~{\rm
  GeV}$, KK neutrino production results in cosmological overclosure. This problem persists even at
$m_\psi \sim 10~{\rm GeV}$ because the
distribution's tail continues to populate the KK ladder with unwanted
neutrinos. Consequently, to keep the neutrino tower subdominant to the
graviton tower in the early universe, we adopt $m_\psi \sim 100~{\rm
  GeV}$. Next, we discuss why this mass scale represents a particularly compelling choice.

Following~\cite{Montero:2025hye}, we assume that $U(1)_{B-L}$ propagates in the bulk, acts
as a global symmetry on the brane, and its brane anomalies are
cancelled by inflow from the bulk. In addition to the three families
of Majorana fermions $\psi$ of $B-L$ charge $q$, the bulk contains a gauge
field and a complex scalar $\Phi$ of charge $-2q$ whose vacuum
expectation value Higgses $B-L$. The renormalizable coupling $\Phi^*
\bar \psi^{c} \psi$ is consistent with the symmetries of the theory and the Majorana mass is then of the
order of the vacuum expectation value $\langle \Phi \rangle $ which would not be unnatural to assume is also of the order $ \sim
\langle H \rangle$.

The gauged $B-L$ dark dimension with $m_\psi \sim 100~{\rm GeV}$ is
fully consistent with current experimental data and standard neutrino
phenomenology. Low-energy neutrino oscillations align with SM predictions, as active-sterile mixing and unitary violations are
suppressed to the negligible level of $\sim 10^{-14}$. Consequently, direct
measurements from KATRIN~\cite{KATRIN:2024cdt} and cosmological observations~\cite{Planck:2018vyg} yield standard,
ordinary results for the absolute neutrino mass scales. The model also comfortably evades all high energy
constraints; Higgs, $Z$, and $W$ boson decays are unaffected, electroweak
precision observables remain unperturbed, and the active-sterile
mixing angle is at least twenty orders of magnitude below the
sensitivity limits of direct collider searches for heavy neutral
leptons~\cite{Abdullahi:2022jlv,ParticleDataGroup:2026mpi}.

In closing, we note that the bulk-coupled Weinberg operator 
\begin{equation}
S \supset \int d^{4}x\int dy\,\delta (y)\left[\frac{(LH)^{2}\Phi
  }{M_{5}^{5/2}}\right] \,,
\end{equation}
is also consistent with the symmetries of the theory. Since $\Phi$ has a volume suppression because it
lives in the bulk, the contribution to the mass of the active neutrino
$m_\nu = m_L - \sum_n m_D^2/m_n $ is negligible,
\begin{equation}
m_L \sim \frac{\langle H \rangle^2 \, \langle \phi
\rangle}{M_{5}^{5/2}\sqrt{R}} \sim \frac{\langle H \rangle^2 \langle
\phi \rangle}{M_5 M_p} \sim 10^{-12}~{\rm eV} \, .
\end{equation}
It is important to note the constrast with pseudo-Dirac neutrinos,
where the lepton number is broken only by $m_L$, yielding a pair of
Majorana states at $m_D \pm m_L/2$~\cite{Beacom:2003eu}.
Solar neutrinos are sensitive to $\delta m^2 = 2 m_D
m_L$ through active-to-sterile disappearance of the states present
in $\nu_e$, and the observed solar fluxes bound $\delta m^2 \alt 1.8
\times 10^{-12}~{\rm eV}^2$~\cite{deGouvea:2009fp,Ansarifard:2022kvy}. Astrophysical neutrinos at IceCube (Mpc
baselines and TeV energies) probe $\delta m^2$ down to about $10^{-18}~{\rm eV}^2$~\cite{Carloni:2022cqz}. While the gauged $B-L$ dark dimension predicts a
continued absence of pseudo-Dirac splitting, IceCube will be
well-positioned to test the model's viability upon a potential
splitting discovery.

\section{The Black Hole Portal}
\label{sec:4}

It has long been recognized that primordial black holes could have
formed in the early universe~\cite{Zeldovich:1967lct,Hawking:1971ei} potentially constituting a
non-negligible fraction of the dark matter~\cite{Chapline:1975ojl}. Within the dark dimension
framework, any such black holes necessarily possess a 5d
nature~\cite{Anchordoqui:2025opy}. Given that the evaporation of these
bulk localized objects 
may provide a viable mechanism to excite the KK towers above the mass
gap (defined here as the mass of the first excited state among the $B-L$ gauge
boson, bulk scalar, and neutrino towers which we continue to denote by $m_\psi$) it is of interest to explore
whether they can serve as a portal to probe the particle spectrum of the dark dimension. The
underlying mechanism is straightforward:  black holes with Hawking temperature $T_H< m_\psi$
emit only the graviton tower, but once the
temperature surpasses the mass threshold,  black holes
begin radiating $(B-L)$ KK gauge bosons, scalars, and neutrinos. The subsequent decay of these KK
states produces a distinct electromagnetic signature, providing a
target for active experimental searches. 

A 5d Schwarzschild black hole is
characterized by its horizon radius $r_h$ and its Hawking temperature $T_H$,
which are inversely proportional to one another, i.e.
\begin{equation}
  T_H \sim \frac{1}{r_h} \, .
\label{TH}
\end{equation}  
By relating the horizon radius to the 5d Planck scale,
\begin{equation}
r_h \sim \frac{1}{M_5} \  \left(\frac{M}{M_5} \right)^{1/2}     \,,
\end{equation}  
the black hole mass $M$ can
be explicitly expressed as a function of its temperature
\begin{equation}
  M(T_H) \sim \frac{M_5^3}{T_H^2} \, .
\label{MTH}
\end{equation}
This implies that a black hole evaporating today with a temperature
above the mass gap $T_{H,{\rm today}} \agt 100~{\rm GeV}$ has $M_{\rm today} \sim 10^{23}~{\rm
  GeV}\sim 1~{\rm g}$. After accounting for greybody factors, the bulk graviton power across all five
polarizations drives the
evaporation rate
\begin{equation}
  \frac{dM}{dt} \sim \alpha \frac{M_5^3}{M} \, ,
\end{equation}
with $\alpha \sim 10^{-3}$~\cite{Cardoso:2005mh}. The lifetime scales as
\begin{equation}
  \tau \sim   \frac{1}{2 \alpha} \ \frac{M^2}{M_5^3}
\end{equation}
and therefore a black hole completing its evaporation today $\tau \sim
t_{\rm U}$ was born with $M_{\rm initial} \sim 10^{33}~{\rm GeV}$ and
an initial temperature $T_{H, {\rm initial}} \sim 1~{\rm MeV}$  
 where $t_U \sim 4 \times 10^{17}~{\rm s}$ is the age of the Universe.\footnote{We note in
  passing that primordial black
  holes could potentially originate from the gravitational collapse of
  cosmic string loops that contract below their Schwarzschild
  radius~\cite{Hawking:1987bn}.  One mechanism for this in the dark dimension, assuming there is a GUT gauge boson, is to use the string with tension $(10^9\ {\rm GeV})^2$ which is needed for its realization~\cite{Heckman:2024trz}.  Another option is to use a brane-localized axion in the dark dimension, with
  symmetry-breaking scale $f_a \sim 10^9~{\rm GeV}$~\cite{Gendler:2024gdo}, the axionic 
  string tension evaluates to $\mu \sim 2 \pi f_a^2$. For loops at
  temperature $T$, the maximum mass of the resulting black hole is
  reached when the loop size exactly equals the horizon scale at the
  moment of collapse $M_{\rm max} \sim \mu/H(T)$~\cite{Anchordoqui:2025opy}. At
  $T_* \sim 1~{\rm GeV}$ this mechanism then yields  $M_{\rm max} \sim 10^{37}~{\rm GeV}$.} These black holes are far
longer lived than 4d black holes of the same mass, which would have evaporated in seconds to days.

\subsection{Bound on Primordial Black Hole Dark Matter}

5d Schwarzschild black holes radiate isotropically into all four spatial
dimensions. The emission probability is uniformly distributed over the
$S^{3}$ hypersurface with line element
$dA=\sin ^{2}\theta \,d\theta \,d\Omega _{2}$, where $\theta \in [0,
\pi]$ is the angle with the dark dimension and $d\Omega_2$ the angular element of $S^2$. Integrating out the brane
angles yields a factor of $4\pi$. The normalized ($\int _{0}^{\pi
}P(\theta )\,d\theta = 1$) probability density function is $P(\theta
)= (2/\pi) \sin ^{2}\theta$. A massless 5d graviton of energy $E$
emitted at angle $\theta$ relative to the dark dimension has a bulk
momentum component $p_5 = E \cos
\theta$ (yielding a 4d KK mass $m_4 = E|\cos\theta|$) and a
brane-localized momentum $E \sin \theta$, resulting in
a 4d Lorentz factor $\gamma = 1/|\cos
\theta|$. The expected value of the fractional rest mass radiated by
the black hole is then
\begin{equation}
\langle m_{4}/E \rangle=\int _{0}^{\pi }|\cos \theta
|\, P(\theta )\,d\theta = \frac{4}{3\pi } \,, 
\end{equation}
and since the total energy fraction is normalized to one, the
average 4d kinetic energy fraction carried away by the emitted KK
gravitons is
\begin{equation}
\langle E_{\rm kin}/E \rangle=1- \langle m_{4}/E\rangle
\simeq 0.58 \,.
\end{equation}
The gravitons of the same KK masses that make up the dark matter tower
were also produced with momenta of order their mass, but have redshifted by a factor $\sim 10^{12}$ since then and are cold; the
gravitons emitted by the black holes are produced today, and have no
time to redshift. The intra-tower cascade does not change this: due to
the small KK-number violation, the decay is governed by tiny mass
splittings, ensuring the daughters closely inherit the parent's
velocity. The black hole evaporation therefore converts
cold dark matter into dark radiation at late times. Now, the effective fraction of dark matter
converted into dark radiation is given by
\begin{equation}
  f_{\rm eff} = \langle E_{\rm kin}/E \rangle \ C_M \ f_{\rm BH} \,,
\end{equation}  
where  $f_{\rm BH}$ is the dark matter fraction in the form of
primordial balck holes and $C_M$ is a dimensionless mass-function dependent scaling factor
of the total mass loss in the evaporating black hole population from redshift $z \sim
10$ to the present.

Planck and BAO data strictly limit dark radiation in the local universe by placing tight
constraints on the suppression of large-scale structure formation. If
dark matter decays into dark radiation, it induces free-streaming and
dark acoustic oscillations that erase
small-scale density perturbations. By combining Planck CMB
lensing with late-time BAO clustering, we can precisely measure the matter
power spectrum and the growth rate of structure. Because these combined datasets
reveal no significant deficit in small-scale power, they severely
restrict the decay rates within the dark sector, ensuring that local
structure matches the standard cold, collisionless dark matter
paradigm. Specifically, no more than 3\% to 6\% of dark matter could
have decayed into radiation over the lifetime of the universe~\cite{Poulin:2016nat,Alvi:2022aam}.

To establish a conservative and model-independent bound that
does not rely on unobservable high-mass or low-mass tails of an
extended distribution, we cast the limit directly in terms of the
total current black hole dark matter fraction, $f_{\rm BH}$. By isolating the population
strictly to the active late-time window, the effective mass conversion
coefficient reduces to its minimal baseline, $C_M =1$. Taking the
kinetic emission fraction of 58\% as the dark radiation contribution, the
Planck and BAO limit $f_{\rm eff} < 0.03$-$0.06$ simplifies directly to $f_{\rm
  BH} < 0.05$-$0.1$, providing a robust, distribution-agnostic upper bound.

In summary, the emission of dark gravitons into the bulk
introduces a highly sensitive correction for primordial black holes exploding today, establishing a new bound on the fractional
abundance of dark matter composed of 5d black holes. This refines the
traditional lower mass threshold, which is typically derived by evaluating the black hole lifetime~\cite{Anchordoqui:2022txe}.

\subsection{Galactic Halo Gamma-Ray Flux}

We now turn to explore potential signals from KK decays into SM
particles. Modeling the Galactic halo with an extended Navarro-Frenk-White (NFW)
profile $\rho_{\rm DM}(r)$~\cite{Navarro:1995iw} yields an all-sky averaged line-of-sight
integral of~\cite{Guepin:2021ljb}
\begin{equation}
{\cal D}_{\rm avg} = \frac{{\cal D}(4\pi)}{4 \pi} \sim 5 \times
10^{22}~{\rm GeV/cm}^2 \, .
\end{equation}
Thus, black holes exploding at a rate
\begin{equation}
\dot{n} \sim \frac{f_{\rm BH}}{t_{\rm U} \, M_{\rm
  initial}} \sim 2.5 \times 10^{-51}\ f_{\rm BH}~{\rm GeV}^{-1} \, {\rm s}^{-1}
\end{equation}
per unit
dark matter mass, distributed like dark matter, produce an
energy-squared weighted flux  of
\begin{eqnarray}
E^2 \frac{dN}{dE} & \sim & \frac{\dot n \, Y(E) \,
  {\cal D}_{\rm avg}}{4 \pi} \nonumber \\
  & \sim &  1 \times 10^{-6} \ f_{\rm
  BH}~{\rm GeV} \, {\rm
  cm}^{-2} \, {\rm s}^{-1} \,  \, {\rm sr}^{-1} \,,
\label{dNdE}
\end{eqnarray}
where  $Y(E) \sim M_{\rm today}$
is the instantaneous KK yield of one black hole decay event and we
have assumed that  black holes with $M_{\rm initial} \sim 10^{33}~{\rm GeV}$
can only constitute a fraction $f_{\rm BH}$ of the total dark matter density.  
The Majorana modes decay through their mixing into $l^\pm W^\mp$, $\nu
Z$, $\nu h$, with 
\begin{equation}
  \Gamma_n \sim \frac{G_F \ m_n^3 \ \theta_n^2}{2 \sqrt{2} \ \pi} \,,
\end{equation}
and asymptotic branching ratios $2:1:1$ (for $m_n < m_h$, $4:1:0$)~\cite{Atre:2009rg}. For,
$\theta_n \sim 10^{-14}$ and $m_n \sim 100~{\rm GeV}$ the lifetime is
seconds to minutes, so the decays are prompt on Galactic scales. Most
of the emitted photons are soft, with hard photons carrying
approximately 18\% of the total energy radiated in the KK mode. If
$m_{B-L} < m_\psi$ the black hole reaches the mass threshold to emit
gauge bosons first. Because the bosons directly decay into fermion
pairs, the energy transferred to photon increases. However, this is
countered by a decrease in the number of available degrees of
freedom. These two effects roughly balance out, yielding a similar
overall outcome.

The predicted gamma-ray diffuse emission can be directly tested against data from the Fermi Large
Area Telescope (Fermi-LAT). By analyzing the spatial and spectral distribution of the
all-sky gamma-ray background, Fermi-LAT searches for a characteristic
anisotropic halo component tracing the underlying NFW dark
matter profile~\cite{Fermi-LAT:2012pls}. Should no signal appear, it can establish an upper
bound on $f_{\rm BH}$ via (\ref{dNdE}).

We end with a speculative observation. A recent analysis of fifteen years of
Fermi-LAT data identifies a statistically significant ($13–19\sigma$)
gamma-ray excess in the Galactic halo,
within the high-latitude window defined by
$|l| \leq 60^\circ$ and $10^\circ \leq |b| \leq
60^\circ$~\cite{Totani:2025fxx}. The
excess is spherically symmetric about the Galactic Center, is detected
consistently in all four Galactic quadrants and in comparisons of
low-latitude regions with the poles, and shows a broad non–power-law
spectrum: the intensity rises from 1~GeV, peaks in the 21~GeV bin, and
drops rapidly, becoming consistent with zero above 200~GeV. The flux
intensity normalized at the Galactic pole reaches
\begin{equation}
  E^2 \frac{dN_\gamma}{dE} \sim 4 \times 10^{-8}~{\rm GeV} \ {\rm cm}^{-2} \
  {\rm s}^{-1} \ {\rm sr}^{-1} 
\label{dNdEg}
\end{equation}
at the 21~GeV peak~\cite{Stenhouse:2026nve}.

The orders of magnitude of (\ref{dNdE}) and (\ref{dNdEg}) are comparable. Because 
the observed radiation excess is only measured at higher galactic
latitudes ($|b| \geq 10^\circ$), we can further explore whether
  a black hole population that does not trace the dark matter
  everywhere, but only in the region probed by the window, could
  produce the same flux using less total mass. For example, if the
  black hole density $\rho_{\rm BH} (r)$ drops off sharply
  at the outer edges of the galaxy according to $\rho_{\rm BH} (r) \propto \rho_{\rm DM} (r) e^{-(r/R_{\rm out})^\beta}$ (with an outer radius $R_{\rm out} =
20~{\rm kpc}$ and a steepness
factor $\beta = 6$), the predicted radiation excess remains virtually
identical. Crucially, this
specific configuration requires only 11\% of the total dark matter
halo mass. If primordial black holes are such a ``biased tracer'' of dark matter,
    similar to material accreted early into the halo would be, the
    resulting average radiation flux increases by an order of
    magnitude for the exact same overall black hole fraction $f_{\rm BH}$. Nonetheless, the absolute
    predictability of this signal remains subject to severe uncertainties spanning
multiple orders of magnitude, driven by galactic modeling, halo profile
morphology, mass function fine-tuning, and high-energy physics parameters. 

\section{Conclusions}
\label{sec:5}

Even though having a bulk neutrino is not a necessary ingredient for the dark dimension scenario, we have seen that its existence could have potentially interesting observable effects.  Not only it fits nicely with the existence of $B-L$ guage symmetry  in the bulk and offer an explanation of the active neutrino masses, but also as we have seen here they can serve as a telescope for the dark dimension that can effectively communicate black hole evaporation in the bulk to the SM fields, similar in spirit to \cite{Anchordoqui:2025opy}.  We have also seen that it puts a lower bound of about 100 GeV on the mass of sterile neutrinos assuming KK modes constitute the dark matter \cite{Gonzalo:2022jac}.

\acknowledgments{We thank Concha Gonzalez-Garcia for discussion and 
Trinity Stenhouse for email correspondence. L.A.A. is supported by the
U.S. National Science Foundation (NSF Grant PHY-2412679). C.V. is
supported in part by a grant from the Simons Foundation (602883) and
the DellaPietra Foundation. 

We leveraged AI tools, specifically Claude and ChatGPT, for computational cross-checks, manually verifying all final equations.}

\bibliography{refs}
\bibliographystyle{JHEP}

\end{document}